\documentclass[]{spie}  
\usepackage[]{graphicx}
\usepackage{booktabs}
\usepackage{subfigure}

\newcommand{\fig}[1]{Fig.~\ref{#1}}

\title{The Optical Design of CHARIS: An Exoplanet IFS for the Subaru Telescope} 
\author{Mary Anne Peters-Limbach\supit{a}, Tyler D. Groff\supit{a},  N. Jeremy Kasdin\supit{a}, Dave Driscoll\supit{b}, Michael Galvin\supit{a}, Allen Foster\supit{c}, Michael A. Carr\supit{a}, Dave LeClerc\supit{b}, Rad Fagan\supit{b}, Michael W. McElwain,\supit{d} Gillian Knapp\supit{a}, Timothy Brandt\supit{a}, Markus Janson\supit{a}, Olivier Guyon\supit{e}, Nemanja Jovanovic\supit{e}, Frantz Martinache\supit{e}, Masahiko Hayashi\supit{e}, Naruhisa Takato\supit{e}
\skiplinehalf
\supit{a}Princeton University, Princeton, NJ, USA \\
\supit{b}L-3 Communications, Integrated Optical Systems -- SSG, Wilmington, MA, USA \\
\supit{c}Pennsylvania State University, University Park, PA, USA \\
\supit{d}Goddard Space Flight Center, Greenbelt, MD, USA \\
\supit{e}Subaru Telescope, National Astronomical Observatory of Japan, Hilo, HI, USA\\
}

\authorinfo{Further author information: (Send correspondence to Mary Anne Peters--Limbach)\\E-mail: mapeters@princeton.edu, Telephone: 1 609 258 0096}
 
\begin{document} 
\maketitle 

\begin{abstract}
High-contrast imaging techniques now make possible both imaging and spectroscopy of planets around nearby stars. We present the optical design for the Coronagraphic High Angular Resolution Imaging Spectrograph (CHARIS), a lenslet-based, cryogenic integral field spectrograph (IFS) for imaging exoplanets on the Subaru telescope. The IFS will provide spectral information for 138$\times$138 spatial elements over a 2.07 arcsec $\times$ 2.07 arcsec field of view (FOV). CHARIS will operate in the near infrared ($\lambda = 1.15 - 2.5 \mu m$) and will feature two spectral resolution modes   of $R\approx18$ (low-res mode) and $R\approx73$ (high-res mode). Taking advantage of the Subaru telescope adaptive optics systems and coronagraphs (AO188 and SCExAO), CHARIS will provide sufficient contrast to obtain spectra of young self-luminous Jupiter-mass exoplanets. CHARIS will undergo CDR in October 2013 and is projected to have first light by the end of 2015. We report here on the current optical design of CHARIS and its unique innovations. 
\end{abstract}


\section{INTRODUCTION}\label{sec:Intro}
\subsection{Background}
Direct imaging coupled with spectroscopy compliments indirect planet detection by detecting exoplanets at large separations and making possible detailed spectral characterization of exoplanets. The exoplanet community has recently begun to image exoplanets including the HR8799 planets\cite{Marois2008HR8799}, $\beta$ Pic b\cite{Lagrange2008A-probable}, LkCa15b\cite{Kraus2011LkCa}, $\kappa$ And b\cite{Carson2012Direct} and GJ 504b\cite{Kuzuhara2013Direct}. The addition of spectroscopy to direct imaging will be particularly useful in characterizing exoplanets that are like Earth and may support life \cite{Kawahara2012Can-Ground-based}.  Integral field spectrographs (IFSs) are well purposed for taking spectra of exoplanets.  

The purpose of an integral field spectrograph (IFS) is to simultaneously image the spectrum of the full two-dimensional field of view. Our IFS accomplishes this by placing a fast lenslet array at the focus of a slow F/\# beam. The lenslet array effectively serves as a two-dimensional array of slits, creating a sparse image at the lenslet focus. This sparsity provides area for each sampled point to be dispersed without overlapping its spectrum onto the adjacent spectra. The trade off between this minimization of cross-contamination (crosstalk), field of view and spectral resolution while still  Nyquist sampling at the shortest wavelength is the challenge in any exoplanet-purposed IFS optical design.  This type of lenslet array-based spectrograph was first implemented in the visible on the TIGER IFS\cite{bacon19953D}. A few mid- to high- spectral resolution IFSs have since been built and are currently in operation on-sky such as OSIRIS\cite{larkin2006osiris}. CHARIS is a low spectral resolution (R = 15 -- 100) IFS similar to Project 1640\cite{Hinkley2010A-New-High}, GPI\cite{macintosh2008gemini, Perrin2010The-Integral} and SPHERE\cite{beuzit2008sphere, Claudi2011Optical} and in that it is designed to image exoplanets. 

\begin{figure}[h]
\centering
\includegraphics[width=0.95\textwidth]{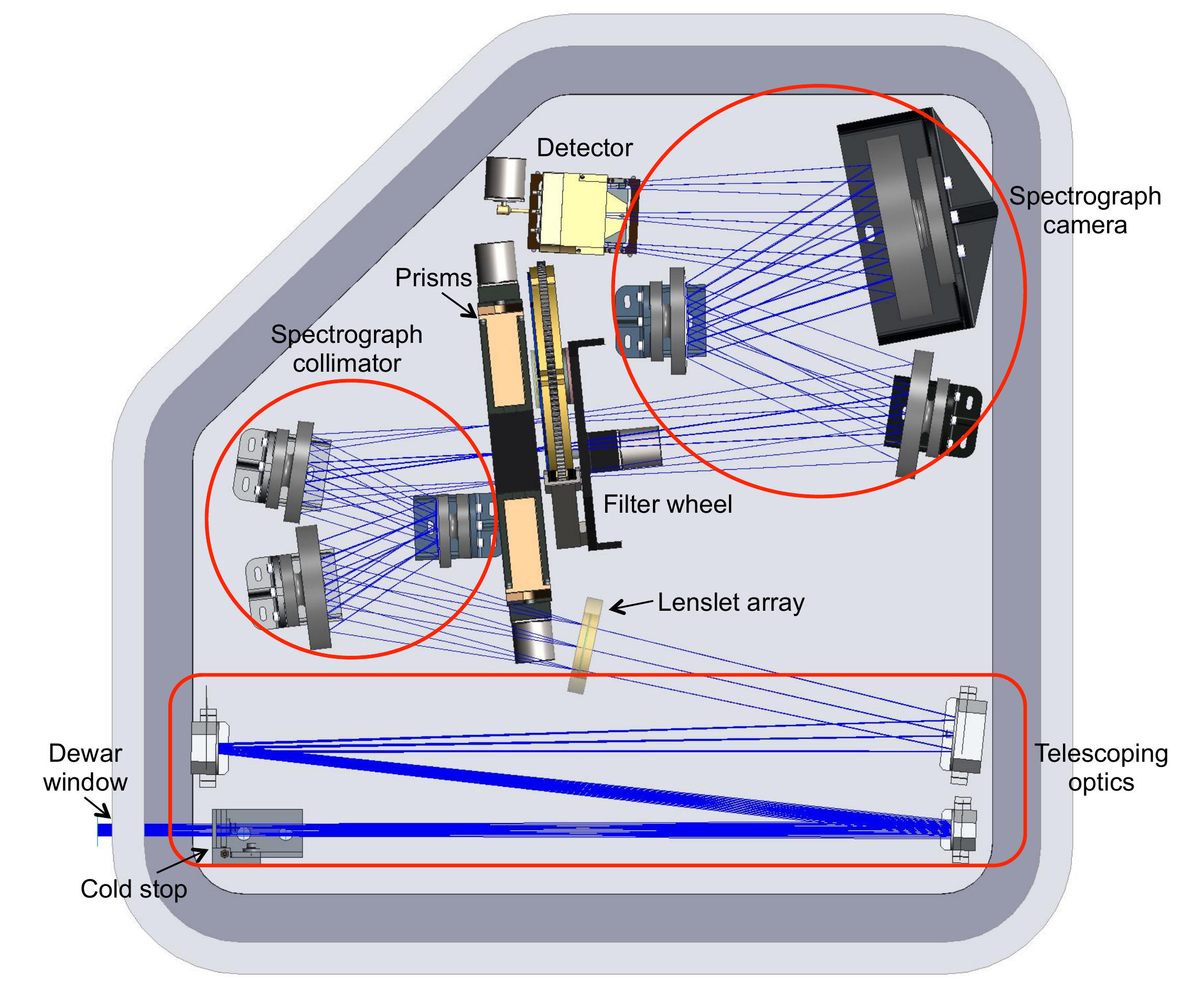}
\caption{Overall Layout of the CHARIS optical assembly, including rays, optomechanics and optical bench. The key optical components discussed in this paper are labeled in this figure and will be referred to and discussed in more detail throughout the paper.}
\label{layout}
\end{figure}

Combined with the AO188\cite{Hayano2010Commissioning, Hayano2008Current, minowa2010performance} and SCExAO\cite{guyon2011wavefront, martinache2011subaru} systems, CHARIS will be the first high-constrast exoplanet-purposed IFS on an 8m class telescope in the northern hemisphere able to achieve a small inner-working angle ($2 \lambda/D$) and high contrasts ($ 10^{-4}-10^{-7} $). CHARIS will provide ``high" resolution ($R>60$) spectra in J-, H-, and K-bands and low resolution spectra ($R\sim18$) across all three bands simultaneously in a $2.07" \times 2.07"$ FOV. In this paper we present the optical design in its current state shown in \fig{layout} (which is 3 months prior to CDR-level). We give equations for calculating the fundamental IFS parameters and walk through the CHARIS optical train.
We will also discuss several of the innovative design choices that are unique to the CHARIS IFS. These includes the detailed design of the lenslet array and our optimization of incident f-number, lenslet f-number and lenslet pitch. We will review the usefulness of lenslet pinholes to minimize crosstalk (discussed previously in Woodgate et al. 2006\cite{Woodgate2006AnIntegral}, Bonfield et al. 2008\cite{Bonfield2008GFP} and Peters et al. 2012\cite{Peters2012Conceptual}). The use of pinholes increases the design's sensitivity to phase errors, so we will also examine the sources of phase errors in CHARIS. This will include an estimation of the residual atmospheric phase errors (post AO correction), which we expect to be the dominant source of phase errors. We will summarize the performance of the spectrograph optics including the reasoning for the ensquared energy specifications and their relation to crosstalk. The CHARIS prism designs and resolutions, including the use of new near-infrared glass materials, are also discussed herein. Finally, we will end with a calculation of the transmission and noise for the end-to-end system including atmosphere, telescope, both AO systems and CHARIS optical train and detector. For a more general overview of the CHARIS instrument, please see Groff et al. 2013\cite{Groff2013}.

\subsection{Optical Design Overview}\label{sec:overview}
CHARIS is an integral field spectrograph designed specifically to search for exoplanets within a 2.07"$\times$2.07" field of view (FOV) around the host star in J-, H-, and K-band. The CHARIS optical assembly is shown in \fig{layout}. CHARIS is provided with a collimated beam from the SCExAO instrument. After entering the CHARIS dewar window, the light comes to a pupil plane which is masked with a cold stop. The telescope optics then create the primary image that is incident on the lenslet array. The lenslet array Nyquist samples the primary image, and creates the sparse image. At the focus of each individual lenslet in the sparse image plane, there is a PSF, referred to as a PSFlet. The sparse image plane is collimated by a three mirror compact (TMC). The light is then dispersed by a prism located at the pupil plane of the collimating TMC and passes through a filter, which defines the bandpass of the spectrum. The dispersed beam is then reimaged by the camera TMC, which focuses the light onto the detector. The spacing between spectra, their length, the image plate scale, lenslet/detector pitch, and the dispersion angle relative to the primary axis of the lenslet spacing are all critical to the design of the instrument and must be carefully chosen to meet the overall instrument requirements. The key optical parameters of CHARIS are given in Table \ref{CHARISopPara}. 

\begin{table}[h]
   \centering
   \begin{tabular}{@{} lrclr @{}} 
        \hline
             \multicolumn{5}{l}{{\it Instrument Optical Parameters}} \\
              Parameter (units) & Value & &Parameter (units)  & Value \\
       \hline
      FOV (arcsec) & 2.07 $\times$ 2.07 & & Plate scale (mas) & 15.0\\
      Spatial meas., $X$ (spaxels) &138 $\times$ 138 && Num. spectral meas.& $12-16$ \\
      $R$ (low-res average) & 17.7 && $R$ (high-res average) & 72.7\\
      Num. detector pixels, $N_{pixels}$ & 2048 $\times$ 2048 && Detector pixel pitch, $q$ ($\mu m$) & 18.0\\
      Incident F/\# at lenslet ($F/\#_{tele}$) & F/420 & & F/\# at detector & F/8.5 \\
      Length of spectrum, $l$ (pixels) & $24-32$ && Gap b/w spectrum, $\delta l$ (pixels) & $1-9$\\
      Width of spectrum, $w$ (pixels) & 2 && Gap b/w spectrum, $\delta w$ (pixels) & 4.6\\
      Mag. from lenslet to detector, $m$ & 1.0625 & & Dia. of Subaru tele., $D$ (m) & 8.2 \\
        \hline
   \end{tabular}
   \caption{CHARIS optical parameters}
   \label{CHARISopPara}
\end{table}

\section{CHARIS Pre-lenslet Optics}\label{sec:PreLens}
The optical design prior to the lenslet array (labeled in \fig{layout} as the `pre-lenslet optics') consists of the dewar window, cold stop and focusing optics. The function of the prelenslet optics is to take the 9mm collimated beam received from SCExAO and focus it onto the lenslet array at F/420. The collimated beam received from SCExAO enters CHARIS through an Infrasil dewar window $\sim$10mm thick and 15mm in diameter. The dewar window will be AR-coated to increase transmission and will have a 30 arcminute wedge on the back surface to decrease ghosting. The beam comes to the pupil plane 20mm inside of the dewar. The cold stop is located directly at the pupil. It is oversized by 0.6mm to accommodate slight misalignments but to still block the maximum amount of unwanted straylight. The central edge of the cold stop is comes to a knife edge 100$\mu m$ thick to minimize diffraction from the stop.

After the coldstop, the telescope optics function together to obtain the required f-number of 420 at the lenslet array. As seen in \fig{layout}, the telescope optics consist of three mirrors, referred to as M1, M2 and M3. The first two mirrors, separated by 600mm, are spheres with radii of curvature 1579mm and -479mm, respectively. M3 is a flat used to fold the beam for instrument volume constraint purposes. The f-number of this design is F/420.2, however the exact f-number of the system is likely to vary by $\pm$5\% due to the extreme sensitivity of the f-number to the distance between M1 and M2. Even  though the optics are spheres, the RMS WFE is only 7nm because the f-number is so slow. Furthermore, at the spatial frequencies corresponding to the exoplanet discovery zone, 1--32$\lambda/D$, the mirrors only have 1nm RMS WFE. This surface quality corresponds to a degradation in Strehl of only about 1\% (i.e. from 90\% to 89\% Strehl).

 \begin{figure}[h]
\centering
\includegraphics[width =  \textwidth]{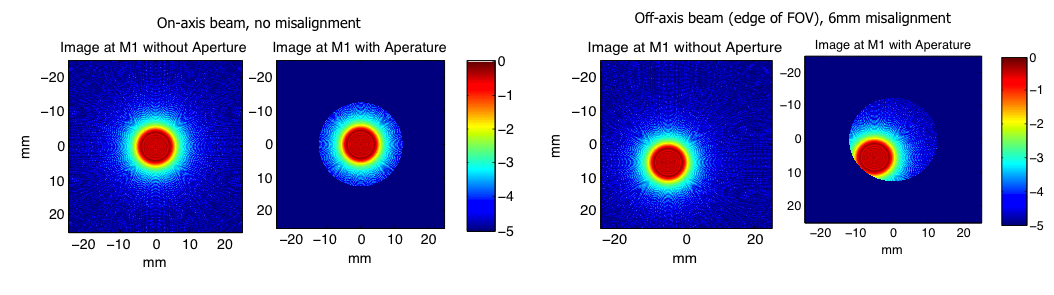}
\caption{This figure illustrates the diffraction at M1 which is calculated by Fresnel propagating the beam from the pupil/cold stop to M1. The first and third figures show the diffraction pattern extended out beyond the mirror aperture to illustrate the amount of light lost, whereas the second and fourth figures show the finite aperture size (25mm in this case) of the M1 mirror. Note that the color bar shows the normalized image intensity and is on a log scale. The left two figures are for an on-axis beam with perfect alignment and the right two are for an off-axis beam at the edge of the field with 6mm of misalignment.}
\label{M1diffLoss}
\end{figure}

To determine the optimum diameter of the telescope mirrors we performed a Fresnel diffraction analysis.  \fig{M1diffLoss} shows an example of this analysis on M1. The throughput loss for the on-axis, perfectly aligned case is 0.2\%, and the loss for the off-axis beam with 6mm misalignment (worst case scenario) is 1.0\%. The actual diameter of M1 is likely to be $\sim$35mm, which will allow for further misalignment on this mirror (with similar throughput numbers) to adjust the tip/tilt of the beam.

\section{Lenslet Array}\label{sec:LensArray}
\subsection{First Order Parameters}\label{Ldesign}

\begin{figure}[h]
\centering
\includegraphics[width = 0.65\textwidth]{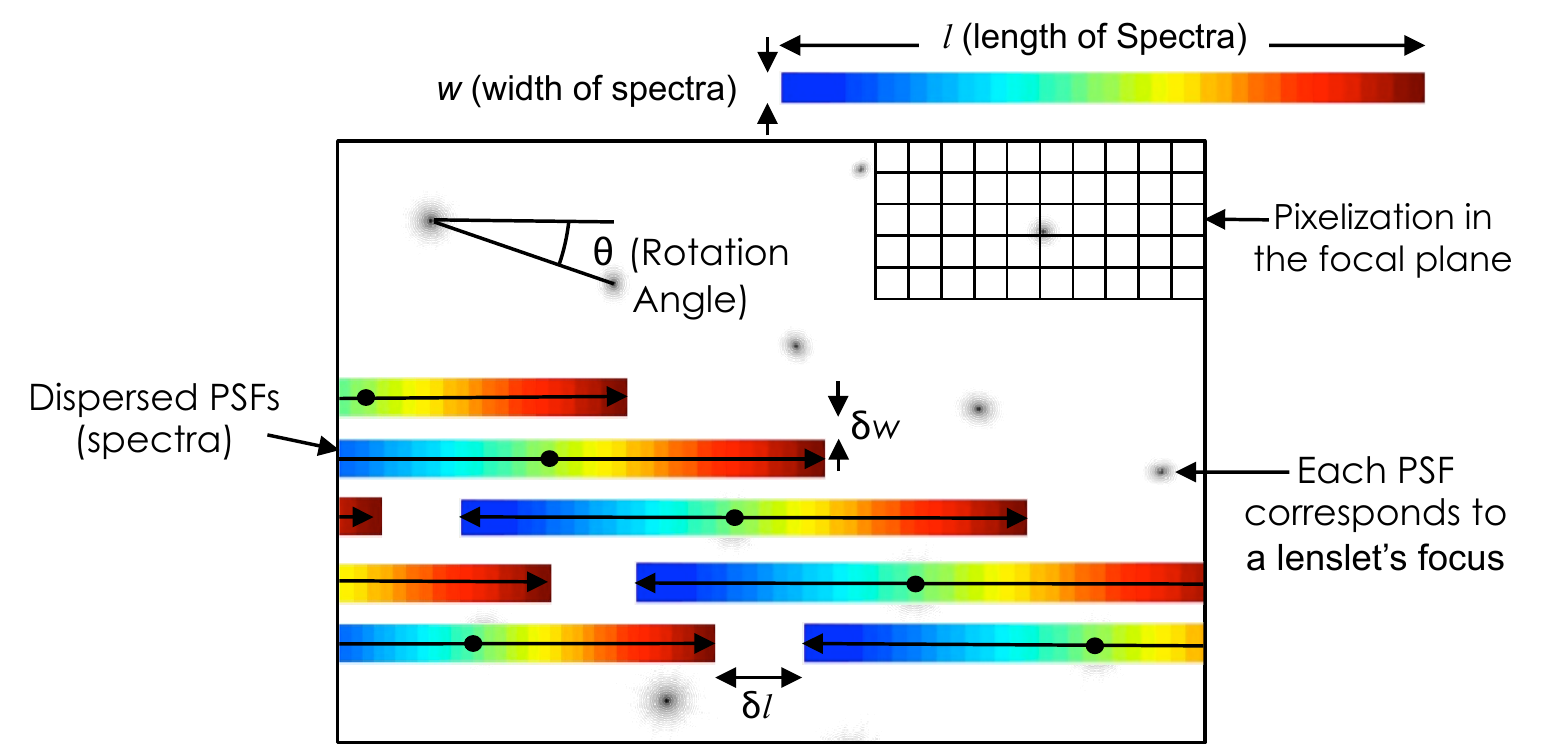}
\caption{This figure shows the detector with several undispersed point spread functions (PSFlets) and dispersed PSFlets (spectra). Each PSFlet in the detector plane (shown here) corresponds to the focus of a single lenslet in the array. This figure also illustrates several of the quantities discussed in the text including spectral length and width, the gap between spectra and the rotation angle of the lenslet with respect to the prism/detector.}
\label{rotation}
\end{figure}

At the focus of the F/420 beam provided by the prelenslet optics is the lenslet array. The lenslet array design was optimized by varying three parameters: the incident f-number of the telescope optics, the lenslet f-number and the pitch of the lenslets. Choosing these three parameters constrains the system to discrete rotation angles (rotation is necessary to avoid overlapping spectra, see \fig{rotation}). These three parameters along with the detector pitch, number of pixels and spectrograph magnification, allow for the calculation of most instrument parameters such as the length and width of the spectra, the gap between the spectra, and the instrument's FOV. In this section we will discuss how we chose the three lenslet array parameters and calculate the key CHARIS parameters.

To design the lenslet array, we first explored the trade space between incident F/\#, lenslet F/\# and lenslet pitch. Based on the trade studies, we were able to conclude a few rules of thumb. In general, increasing the PSFlets separation by increasing pitch while holding the lenslet f-number fixed will decrease lenslet crosstalk. Faster f-number lenslets will have a smaller diffraction pattern (and hence less crosstalk), but the spectrograph's optics must also be increasingly fast as the lenslet f-number speeds up. Faster spectrograph optics typically correlate with a more costly design and increased optical wavefront error (WFE). Finally, a slow incident beam decreases the phase variation across the lenslets, which minimizes the PSFlet translation in the lenslet focal plane. In other words, we require that the lenslet f-number be much faster than the incident f-number so that the lenslets effectively see collimated light. This design makes the depth of focus of the telescoping optics is on the order of the lenslet focal length.

This trade study drove us to a lenslet design with a F/420 incident beam and F/8 lenslets with a  250$\mu m$ pitch. These f-numbers are slower than those typical of existing high-constast imaging IFSs. We found that F/8 was the optimum lenslet f-number in the trade space of spectrograph optics cost and performance (driving us to a slower lenslet f-number) verses minimizing crosstalk (which drove the faster f-number). The F/8 spectrograph was slow enough to allow for the use of reflective optics which will decrease the chromaticity and increase the throughput of the system, minimize the size of the collimated pupil, and reduce the complexity of the optical figure in the collimator and camera. The slower f-number should also make these optics easier to design, build and assemble while still meeting our wavefront error (WFE) specifications. An incident telescope f-number of F/420 was chosen because of concern for diffraction effects from a slower telescope (which would require larger diameter focusing optics, see \fig{M1diffLoss}) and because the depth of focus ($ = 1.18$mm) of the telescope optics is of order the lenslet's focal length. This f-number also corresponds approximately to a 1:1 imaging system between the lenslet and detector. Minimizing spectrograph magnification (i.e. slower incident f-numbers) decreases crosstalk because the PSF is magnified less. To achieve Nyquist sampling, we chose the lenslet pitch to be 250$\mu m$ which gave a plate scale of 
\begin{equation} \label{eq:platescaleIS}
	\phi = \frac{p}{(F/\#_{tele})(D)} = 15.0\textrm{ mas/lenslet},
\end{equation}
where the variables in Eq. \ref{eq:platescaleIS} -- \ref{eq:xSpatial} are defined in Tables \ref{CHARISopPara} or \ref{CHARISLensPara}. This plate scale corresponds to Nyquist sampling at $\lambda = 1.19\mu m$. Note that the lenslet array defines the spatial sampling in the system and the detector only defines the spectral sampling. The rotation angle of the lenslet array is given by 
\begin{equation} \label{eq:thetaROT}
	\theta = \textrm{arctan}\left(\frac{n_y}{n_x}\right) = 26.565^{\circ}, n_y = 1
\end{equation}
where $n_x$ = 2 and $n_y$ = 1 are the spaxels\footnote{A spaxel is one spatial element, corresponding to a singe lenslet, in an IFS.} along the x and y axis, respectively, that give the rotation angle of the lenslet relative to the unclocked position and where the angle of lenslet clocking is measured relative to the dispersion direction. Integer multiples of lenslets ($n_x$ and $n_y$) must be chosen to obtain the rotation angle. Otherwise, the spectra will not line up in a row as shown in \fig{rotation}, but will rather be offset, possibly overlapping and a poor use of detector real estate. We note that rotating the lenslet has the same effect as rotating the prism and detector. For CHARIS, we choose to rotate the lenslet rather than the prism so that (1) only the lenslet needs to be tilted rather than tilting multiple prisms and the detector, (2) it is easier to calibrate the location of the spectra in the absence of the prism, and (3) the camera is hard mounted rather than a moving part.

Once a rotation angle is chosen the length and width between spectra can be calculated using the rotation angle, $\theta$, the magnification of the system, $m$, the lenslet pitch, $p$, and the pixel pitch, $q$. The center-to-center horizontal separation between spectra, or equivalently the length of the spectra, $l$,  plus the horizontal gap between two adjacent spectra, $\delta l$, is given by 
\begin{equation} \label{eq:lengthSpec}
	l + \delta l = \frac{n_xpm}{q\textrm{ cos}\theta} = 33.0 \textrm{ pixels}.
\end{equation}
The center-to-center vertical separation between spectra, or equivalently, the width of the spectra, $w$, plus the vertical gap between adjacent spectra, $\delta w$, is given by 
\begin{equation} \label{eq:widthSpec}
	w + \delta w = \frac{n_ypm\textrm{ sin}\theta}{q} = 6.6 \textrm{ pixels}.
\end{equation}
The number of spaxels, in the x or y direction is the same and is given by
\begin{equation} \label{eq:xSpatial}
	X = \frac{qN_{pixels}}{pm} = 138 \textrm{ spaxels}.
\end{equation}
Finally, the FOV is simply given by the number of spaxels times the plate scale:
\begin{equation} \label{eq:FOV}
	FOV = (X)(\phi) = 2.07  \textrm{ arcseconds}.
\end{equation}
A comprehensive list of lenslet parameters is given in Table \ref{CHARISLensPara}.

\begin{table}[h]
   \centering
   \begin{tabular}{@{} lrclr @{}} 
        \hline
             \multicolumn{5}{l}{{\it Lenslet Parameters}} \\
              Parameter (units)  & Value & &Parameter (units)  & Value \\
       \hline
      Lenslet Pitch, $p$ ($\mu m$) &  250 & & F-number at $\lambda$ = 2.4$\mu m$ & F/8 \\
      Clear Aperture (mm$\times$mm) & 36.7 $\times$ 36.7 & & Thickness (mm) & 2.577 \\
      Material & Infrasil & & Operational temp. (K) & 50 - 300 \\
      Lenslet Shape & Square & &  Surface shape & Spherical \\
      Fill factor & 96\% & & Rotation Angle, $\theta$ (degrees) & 26.565 \\
      Gap between lenslets ($\mu m$) & 5 & & Radius of Curvature (mm) & $0.864$\\
      EFL$^*$ at $\lambda$ = $1.15\mu m$ (mm) & $1.946$ & & EFL at $\lambda$ = $2.4\mu m$ (mm) & $2$\\
   \hline
   \end{tabular}
   \caption{CHARIS lenslet parameters ($^*$EFL = effective focal length)}
   \label{CHARISLensPara}
\end{table}

\subsection{Detailed Design}\label{lensDD}

The lenslet array Nyquist samples the incident point spread function (PSF) and takes each sampled piece of the PSF and focuses it down into a PSFlet concentrating the light by more than 10 fold and creating a sparse image plane, which is necessary to make room for interleaving the spectra on the detector.  The lenslet array specifications are listed in Table \ref{CHARISLensPara}. We are baselining square lenslets because of the high fill factor ($>$96\%) relative to circular lenslets. This high fill factor is another advantage of the slow incident f-number/large lenslet pitch solution: the gap between lenslets is a fixed size, so by having larger lenslets we lose less light between the gaps. 

\begin{figure}[h]
\centering
\includegraphics[width = 0.7\textwidth]{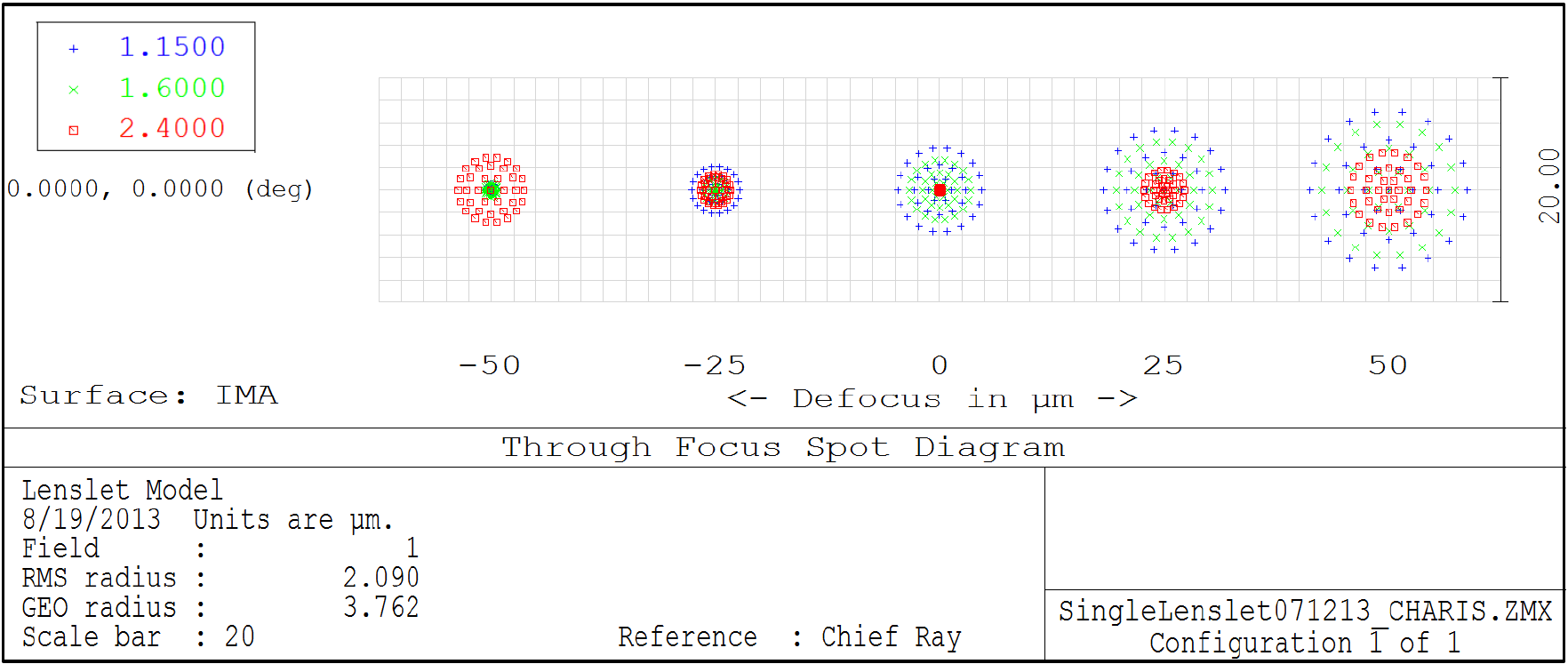}
\caption{Geometrical spot diagrams through the focus of the lenslet array. Note the chromatic defocus.}
\label{throughFocus}
\end{figure}

The lenslet array is the only powered refractive optic in the IFS, and therefore is the primary source of chromaticity. The difference between the focus location of the reddest (2.4$\mu m$) vs. bluest (1.15$\mu m$) wavelength is 54$\mu m$ (see \fig{throughFocus}). In the absence of chromatic focus shown in \fig{throughFocus}, the size of the PSFlets is dominated by diffraction rather than geometrical aberrations and thus the diffraction core at the longer wavelengths is much larger in diameter than at the shortest wavelengths.  We optimize the focus at the longer wavelength to make the full-width half-max of the PSFlet similar at all wavelengths. The geometrical defocus and diffraction give a $\sim$19.2$\mu m$ diffraction core diameter at $\lambda$ = 1.15$\mu m$, which is still smaller than the diffraction core (23.4$\mu m$ at $\lambda$ = 2.4$\mu m$) of the longest wavelength.

 \begin{figure}[h]
\centering
\includegraphics[width = 0.75\textwidth]{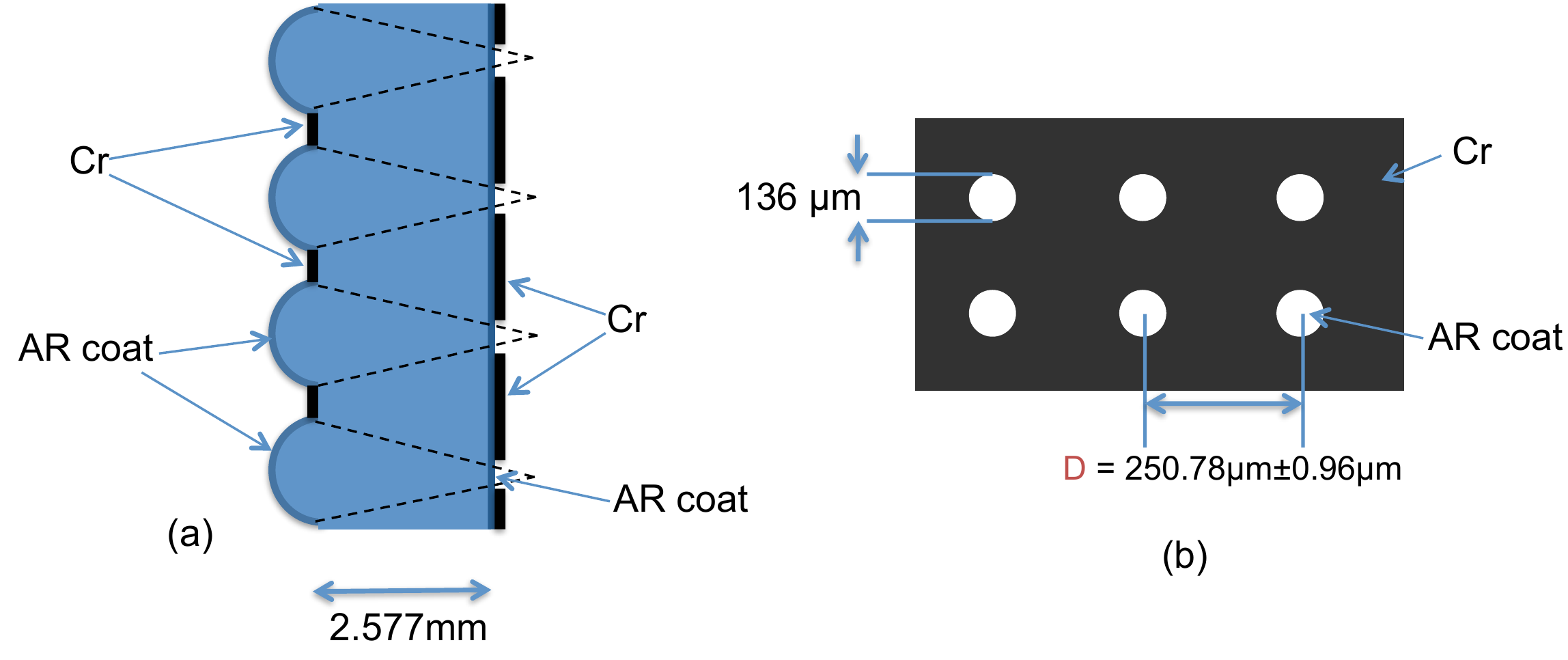}
\caption{(a) Chrome (Cr) and AR coating for the front and back surface of the lenslet array. Note that lenslet focus is chromatic so not all PSFlets come to focus on the back surface of the lenslet. (b) Detailed drawing of the back surface Cr and AR coat, including the spacing between pinholes, $D$, which is slightly larger than the pitch of the pinholes due to the system being non-telecentric. }
\label{lensletCoat}
\end{figure}

Crosstalk is the leakage of the diffracted light from one spectrum into the adjacent spectra. If not dealt with properly, crosstalk will be the dominate source of noise in a high-constrast imaging IFS. To mitigate crosstalk, we have a large separation between adjacent spectra and are using pinholes to block the PSFlet diffraction wings on the back side of the lenslet. Pinholes were implemented for the first time in the GFP-IFS\cite{Bonfield2008GFP}. Our design should suppress crosstalk to $\sim0.1\%$. The details of crosstalk are discussed more in Peters et al. 2012\cite{Peters2012Conceptual} and Groff et al. 2013\cite{Groff2013}. The baseline lenslet design includes pinholes on the back surface of the lenslet array. A chrome coating will be used for the pinholes and placed in the 5$\mu m$ gaps between the lenslets on the front surface. The optic will be also be AR coated to increase transmission and decrease ghosting. A schematic of the coating scheme is shown in \fig{lensletCoat}. The presences of pinholes on the back surface of the lenslet array lead to a stringent phase error requirement which must be carefully calculated and understood. The phase errors are the subject of the next section, \S \ref{sec:phase}. 

\subsection{Phase Errors}\label{sec:phase}
\begin{figure}[h]
\centering
\includegraphics[width = 0.82\textwidth]{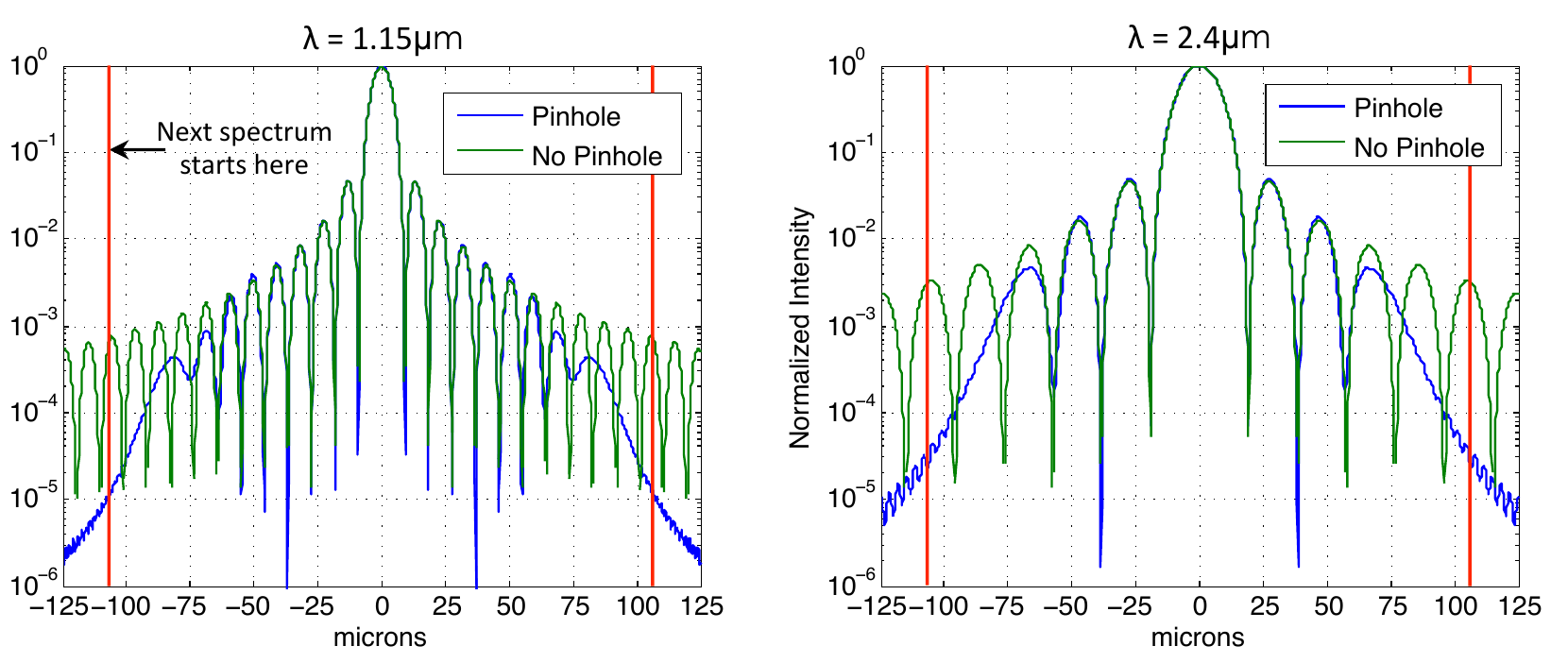}
\caption{Simulated cross-sections at the detector with and without pinholes at the shortest (1.15$\mu$m) and longest (2.4$\mu$m) CHARIS wavelengths. The green line shows the PSFlet without a pinhole. Note that for this case, the crosstalk at the start of the next spectrum (indicated by the red lines) is $\sim$10$^{-3}$ whereas with the pinhole in place (blue line) the crosstalk is suppressed well below 10$^{-4}$ at the start of the adjacent spectrum. Simulation assumes perfect reimaging optics. }
\label{PSFcrossSec}
\end{figure}

Although the pinholes successfully suppress the crosstalk (see \fig{PSFcrossSec}), they can also vignette the beam or completely block the light going through the system if the phase errors on the lenslet are 
severe enough. The lenslet array in a IFS works just like a Shack-Hartmann wavefront sensor, and thus phase errors cause x--y translations in the plane of PSFlets and pinholes. A complete estimate of the phase error due to atmospheric residuals, optical WFE and non-telecentricity is essential to ensure the light is not vignetted by the pinhole array. This section will consist of two parts. First we estimate the required phase error tolerance based on the lenslet pinhole diameter. In obtaining this estimate, we will explore pinhole size and position trades to determine which design allows the maximum phase error while still suppressing the crosstalk to the required levels. Predictable static phase errors (such as non-telecentricity) will be taken into account in this section and incorporated into the pinhole design, while other static phase errors that are not predictable prior to building the lenslet array (such as alignment error) will be taken into account in the allocation of the phase error budget. The second part of this section will estimate the total (non-static) RMS phase error which depends on optical WFE due to all errors prior to the lenslet and residual atmospheric WFE not corrected by the AO systems.

\begin{figure}[h]
\centering
\includegraphics[width = 0.65\textwidth]{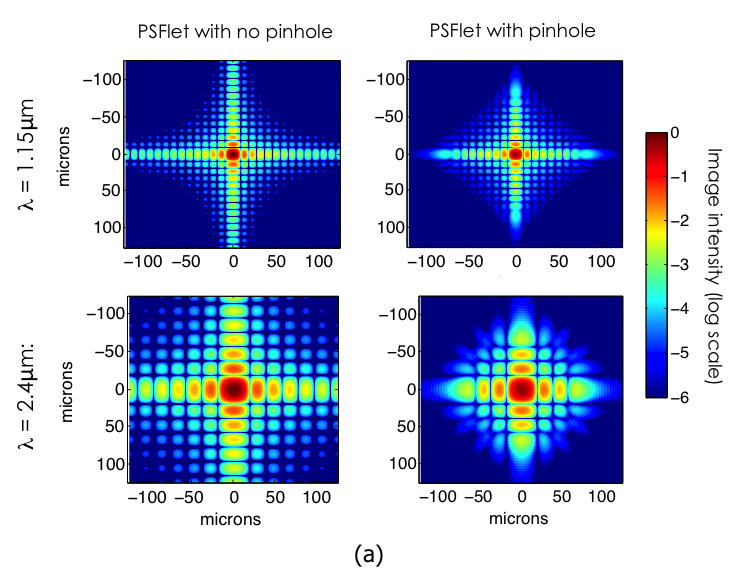}
\caption{(a) PSFlets at the lenslet focus with and without pinholes at the shortest (1.15$\mu$m) and longest (2.4$\mu$m) CHARIS wavelengths. The adjacent spectra starts at $\sim$125$\mu$m from the core. Pinholes and PSFlets are separated by 200$\mu m$ to simulate chromatic defocus.}
\label{PSFlets11502400}
\end{figure}

First we explore the lenslet pinhole phase error sensitivity. Because of chromatic defocus, not all the PSFlets come to focus in the same plane as the pinholes. When we model the sensitivity to phase errors, we use Fresnel propagation code in MATLAB that propagates the light from the front surface of the lenslet array to the PSFlet plane and allows for the PSFlet to be in a different plane than the pinhole to account for chromatic defocus. Simulated phase errors can be added prior to the propagation. The phase error tolerance will be directly proportional to the size of the pinhole behind the lenslet array. Therefore we want to make the pinhole as large as possible. The center-to-center spacing ($w + \delta w = 6.6$ pixels from Eq. \ref{eq:widthSpec}) minus one pixel is the maximum radius the pinhole can be before the diffraction interferes with the adjacent spectra. Thus, we choose a pinhole of radius 4 pixels in the detector plane which corresponds to a 136$\mu m$ pinhole on the back surface of the lenslet array (but is magnified to 144$\mu$m on the detector). Note that the pinhole is undersized to allow for post-lenslet WFEs which will broaden the PSF.  \fig{PSFcrossSec} shows a cross section through the spikes of the PSF in the focal plane of the lenslet array with and without a pinhole. The adjacent spectra is 105$\mu$m away from the peak of the PSF. The goal is to suppress the wings below 10$^{-3}$, which is successfully accomplished by the pinholes with margin for optical WFE.\fig{PSFlets11502400}a illustrates the same information as \fig{PSFcrossSec}, but provides a spatial intensity map rather than a cross section. 

\begin{figure}[h]
\centering
\includegraphics[width = 0.99\textwidth]{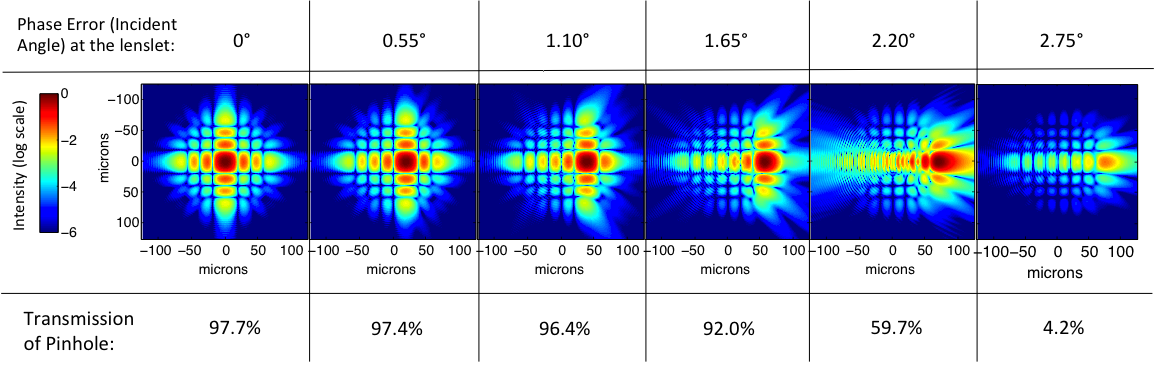}
\caption{Simulated effects of phase error causing misalignment of PSFlet on the pinhole. The total allowed phase error on the lenslet is 1.5$^{\circ}$ (2.73$\lambda$/D). Test wavelength is $\lambda = 2.4\mu m$.}
\label{PinholeMisalign}
\end{figure}

\fig{PinholeMisalign} shows the how the PSFlet changes with varying amounts of phase error (0$^{\circ}-$2.75$^{\circ}$) at the lenslet. The transmission through the pinhole decreases by a small (1.3\%) amount for misalignments of $0-1.1^{\circ}$, but then drops to 92.0\% with a 1.65$^{\circ}$ of misalignment and continues to drop rapidly with larger misalignments. Based on this analysis the total acceptable phase error allowed on the lenslet array is 1.5$^{\circ}$ of misalignment, which corresponds to a 94.0\% transmission through the pinhole (or 3.7\% less than the perfectly aligned pinhole). 

The maximum allowable phase error of 1.5$^{\circ}$ is allocated to various sources. The first allocation is to the actual placement of the pinholes on the back side of the lenslet array. The beam incident on the pinhole is not telecentric, which means that the beam is not perpendicular to the lenslet array off-axis. The incident angle of the chief ray increases by $\phi$ = 0.0224$^{\circ}$/lenslet. This offset is accounted when depositing the lenslet array pinholes to within the manufacture's ability to place the pinhole in the desired location. The corresponding positional shift of the PSFlet per 250$\mu$m (the size of one lenslet) is $0.7814\mu m$/lenslet. The pinhole can be placed in the designated position within $\pm$1$\mu m$ which corresponds to a phase error of $\pm$0.03$^{\circ}$. The phase error budgets are listed in Table \ref{PhaseErrorAlloc}. 

 \begin{table}[h]
   \centering
   \begin{tabular}{@{} lr @{}} 
   \\
        \hline
        \hline
             Parameter & Budgeted Tolerance \\
       \hline
       	Manufacturing budget & 0.03$^{\circ}$ \\
	Alignment budget &  0.05$^{\circ}$\\
	Stability budget & 0.19$^{\circ}$ \\
	Optical WFE budget & 0.02$^{\circ}$\\
	Atmopsheric WFE budget & 1.21$^{\circ}$\\
	\hline
	Total Sum of Tolerances & 1.50$^{\circ}$ \\
   \hline
   \hline
   \end{tabular}
   \caption{Allocation of phase error budget on the lenslet array.}
   \label{PhaseErrorAlloc}
\end{table}

\begin{figure}[h]
\centering
\includegraphics[width = 0.8\textwidth]{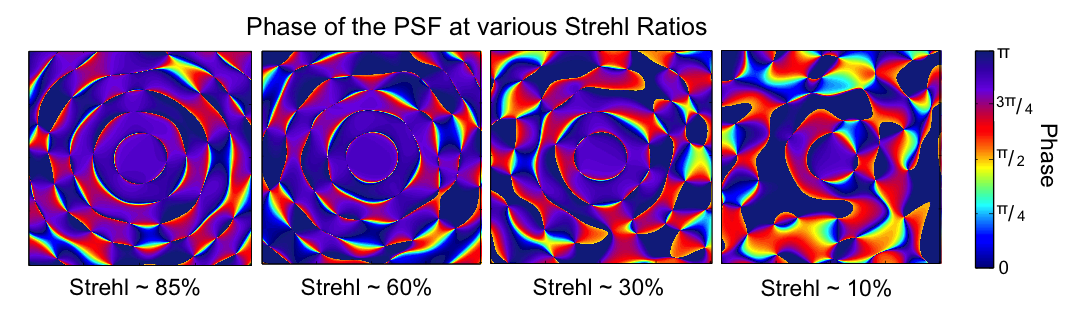}
\caption{Simulated phase maps of the PSF incident on the lenslet array at various input Strehl ratios based on the residual atmospheric WFE after AO correction. Note that the phase in the core of the Airy pattern remains fairly unchanged for Strehls $\geq$ 30\%, and thus causes minimal translation of the PSFlet behind the lenslet array.}
\label{phaseMaps}
\end{figure}

After allocating the phase budget to known sources of phase error, 1.21$^{\circ}$ of phase error remains for atmospheric WFE. To date, the exact atmospheric residual WFE map is unknown. The actual residual WFE maps after AO correction will be measured over the next few months by the SCExAO team and then used to estimate the atmospheric phase error at the lenslet. Until that time, we are using simulated phase maps to estimate phase error. We find that for Strehl ratios as low as 30\% the phase errors in the core of the PSF remain flat and relatively unaffected. The degree to which this statement holds depends on how well the lower order aberrations are corrected, assuming that the lowest order aberrations correspond to the outer scale of atmospheric turbulence. \fig{phaseMaps} shows the phase out to the first four rings of  of the PSF at four different Strehl ratios.

The core of the planet's PSF will be the primary source of photons used for planet detection. It will only be deviated by $\pm0.38^{\circ}$, based on this analysis and the phase map at 30\% Strehl in \fig{phaseMaps} which represents the worst case scenario expected during CHARIS operations. Note that this calculation assumed the lower order aberrations were corrected by the AO system in such a manor that the equivalent outer scale of turbulence in the atmosphere was 1m. If lower order aberrations are present, the results of this simulation will change. The actual residual WFE map to be provided by SCExAO will be used to verify this result. Furthermore, we note that the pinhole itself seems to disrupt the phase error and causes further changes in the translation of the PSFlet. If the location of this pinhole was to change, this would again modify our results.  We also note that the Airy rings outside the central core show variations in phase even at the highest Strehl ratios, and thus the phase errors in the wings of the PSF are expected to be larger that in the core. We intend to investigate the consequences of phase errors more in future papers. 

\section{Spectrograph Design}\label{SpectroDes}
\subsection{Camera \& Collimator Optics}
The spectrograph optics, which include a collimator and camera three mirror compact (TMC) both shown in \fig{layout} surround the prism and the filter. The purpose of these optics is to collimate the light after it exits the lenslet array so it can be dispersed, and then refocused onto the detector. The spectrograph relay optics are being designed and built by L-3 Communications SSG. The basic design parameters for the L-3 optical design are given in Table \ref{L-3DesPara}.

\begin{table}[h]
   \centering
   \begin{tabular}{@{} lrclr @{}} 
        \hline
              Parameter (units) &  Value & & Parameter (units) &  Value \\
       \hline
      Input field size (mm) & 34.7$\times$34.7 & & Image size (mm) & 36.8$\times$37.1\\
      Input field f-number & F/8 & & Image space f-number& F/8.5 \\
      Input pupil distance (mm) & -1133 & &  Magnification & 1.0625\\
       \hline
   \end{tabular}
   \caption{Basic design parameters of the L-3 spectrograph optics.}
   \label{L-3DesPara}
\end{table}

The collimating TMC consists of two parabolic mirrors and one ellipsoid. The camera optics include one parabolic mirror, one ellipsoid and one weak oblate spheroid. The mirror designs were chosen to be testable in a straightforward way to reduce manufacturing risks and cost. The paraboloids are point-testable conics with flat mirror and retroreflecting mirror, the ellipsoids are point-testable conics requiring only a retro-reflection mirror, and the weak oblate spheroid requires only a center of curvature test with simple glass plate null optic. The mirror diameters are comfortably oversized to minimize diffraction and crosstalk, allow for straightforward polishing, and reduce sensitivity to the cryo-environment. The input field size and f-number must match that of the lenslet array. This magnification gives an image space f-number of F/8.5 and an image size corresponding to the the H2RG detector area. Note that the magnification is sightly different in the x- and y-axes and there is $\sim$1\% of distortion in the system. The presence of these aberrations allowed for the design to maximize ensquared energy and minimize other less appealing aberrations. The distortion will be removed in post processing down to the 0.1\% level. We emphasize that because this is a reflective design the system performance changes minimally with wavelength. The use of reflective optics and the corresponding achromaticity of CHARIS is one of the unique and beneficial features of this high-contrast IFS.

The ensquared energy specifications for the L-3 optics were based on the values required to minimize crosstalk to an acceptable level (which drives the ensquared energy value $\pm$5 pixels or $\pm$90$\mu m$ from the center of the PSF), and to undersample or Nyquist sample (which defines the required ensquared energy at $\pm$1 pixel or $\pm$18$\mu m$ from the center of the PSF). Ideally we want the value at 5 pixels to be as small as possible to minimize crosstalk. It turns out the ensquared energy of the spectrograph optics depends almost solely on the f-number (and hence diffraction) in the system as opposed to geometrical WFEs. To calculate the crosstalk for the end-to-end system, both the spectrograph PSF and lenslet PSFlets (which are shown in \fig{PSFcrossSec}) need to be taken into account.  A list of the ensquared energy values for the spectrograph optics is given in Table \ref{L3EE}. 

\begin{table}[h]
   \centering
   \begin{tabular}{@{} lcc||lcc @{}} 
        \hline
               & Nominal  & Allowable  & &   Nominal  & Allowable    \\
                \hspace{2mm}Parameter &  Value &  Range   & Parameter &  Value &  Range \\
       \hline
       \multicolumn{2}{l}{Ensquared energy in 36$\times$36 $\mu m^2$ area:} & & \multicolumn{2}{l}{Ensquared energy in 180$\times$180 $\mu m^2$ area:}\\
         \hspace{4mm}$\lambda$ = 1.1$\mu m$ & 75\% & $>$50\%  &   \hspace{4mm}$\lambda$ = 1.1$\mu m$ & 98.0\% & $>$97\%\\
         \hspace{4mm}$\lambda$ = 1.65$\mu m$ & 71\% & $>$50\%  &  \hspace{4mm}$\lambda$ = 1.65$\mu m$ & 96.5\% & $>$95\%  \\
         \hspace{4mm}$\lambda$ = 2.4$\mu m$ & 64\% & $>$50\%   & \hspace{4mm}$\lambda$ = 2.4$\mu m$ & 94\% & $>$94\%    \\
   \hline
   \end{tabular}
   \caption{Ensquared Energy requirements for the L-3 spectrograph optics.}
   \label{L3EE}
\end{table}

\begin{figure}[h]
\centering
\includegraphics[width = 0.99\textwidth]{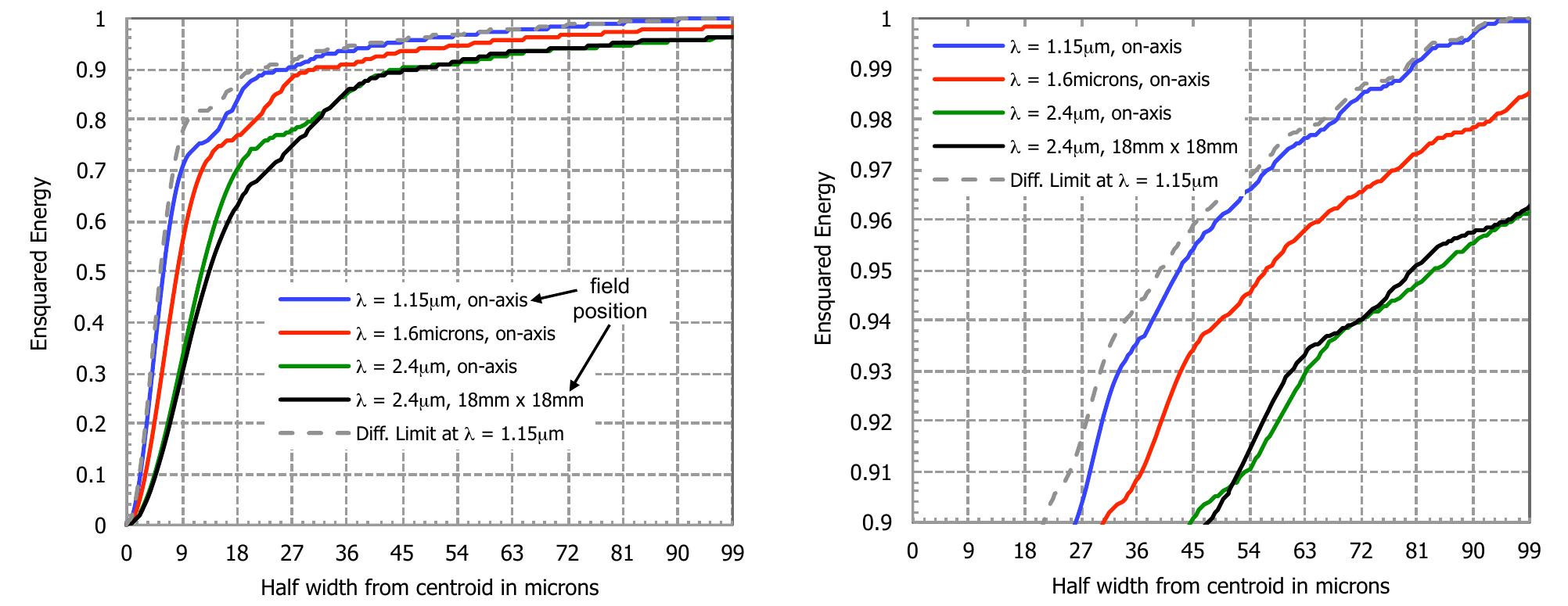}
\caption{Ensquared energy of the spectrograph design at 1.15$\mu m$, 1.6$\mu m$ and 2.4$\mu m$. Left: Zoomed-out version to show the energy distribution from 0\%-100\%. Note that the minimum ensquared energy occurs off-axis at $\lambda$ = 2.4$\mu$m spot. The ensquared energy at the shortest wavelength is $>$80\% on-axis. The 1.15$\mu$m diffraction limit is shown for comparison. Right: Zoomed-in version of the same figure as above to show the ensquared energy at large distance from the centroid. The minimum ensquared energy here is 94\%. Note that the half-width of 18$\mu m$ corresponds to the energy ensquared in a 36$\times$36 $\mu m^2$ area and 90$\mu m$ corresponds to the 180$\times$180 $\mu m^2$ area. }
\label{EE_spectro}
\end{figure}

The ensquared energy plots are shown in \fig{EE_spectro} for the baseline design at 1.15$\mu m$, 1.6$\mu m$ and 2.4$\mu m$. For each wavelength, the ensquared energy values are shown for one or two field locations. The diffraction limit is also shown for the shortest wavelength to allow for comparison between geometrical and diffraction limited performance.

In order to spectrally Nyquist sample the ensquared energy values in the  36$\times$36 $\mu m^2$ area should be 50\%. However, in this case we prefer to undersample by a small amount to increase the amount of light in the $2\times2$ pixel area allotted to recovering a spectral signal. From Table \ref{L3EE} we see that the ensquared energy values are higher than 50\%, however the actual ensquared energy values will decrease further beyond the values given here based on alignment, manufacturing and chromatic defocus from the lenslet array. In Table \ref{L3EE}, we see that the highest ensquared energy values are at the shortest wavelength. Thus, in order to avoid degrading the PSF from the L-3 optics at $\lambda = 2.4\mu m$ further, we optimize the lenslet PSF for the longest wavelength - i.e. the lenslet is focus is optimized for $2.4\mu m$ light and all other wavelengths suffer chromatic defocus as discussed in \S \ref{lensDD}. The chromatic defocus of the lenslet array will drive the bluer wavelengths to have lower ensquared energy values than those quoted for the reddest wavelengths. 

\subsection{Dispersion}\label{sec:Prism}
\begin{figure}[h] 
\centering
\includegraphics[width = .68\textwidth]{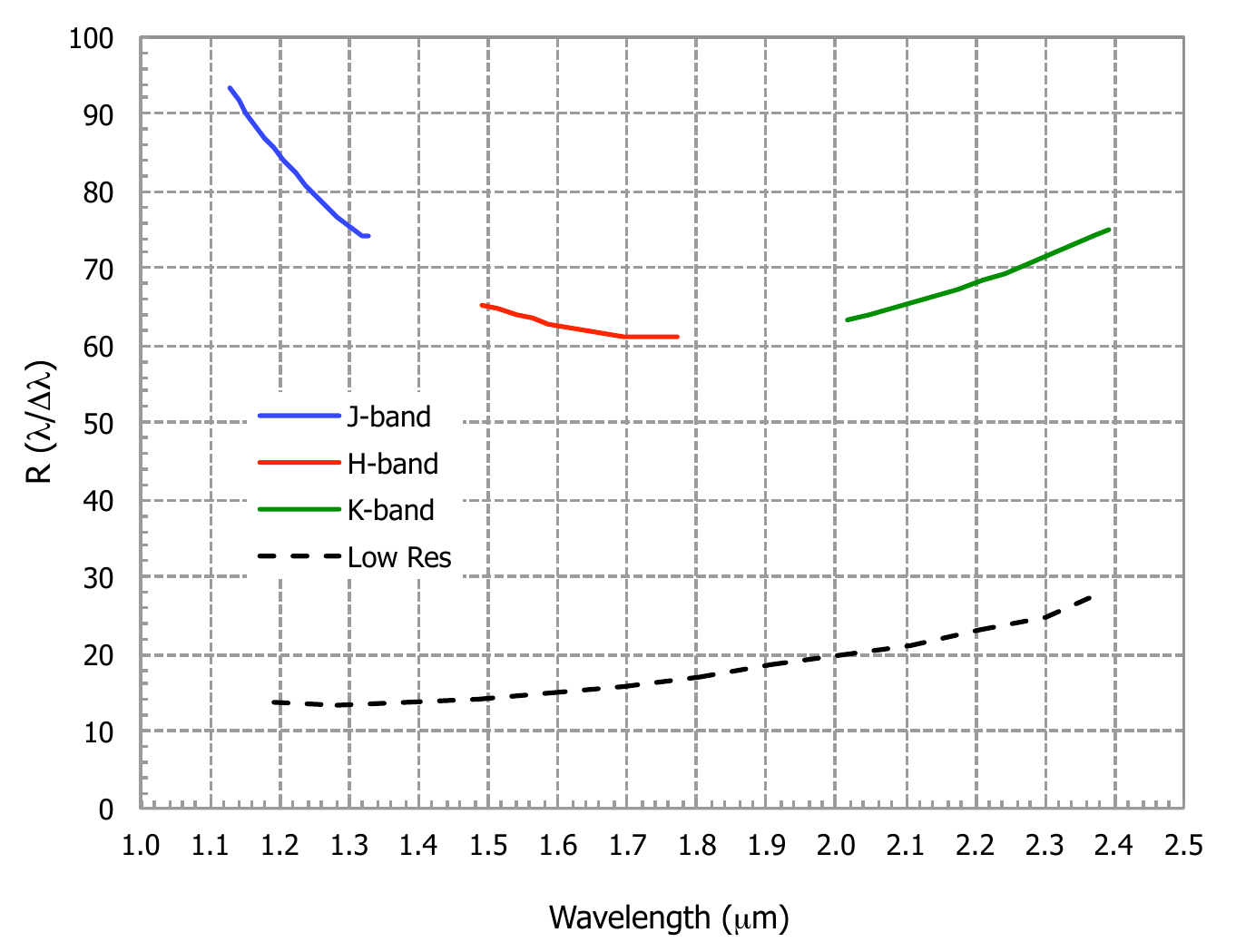}
\caption{Resolution ($\lambda / \Delta \lambda$) as a function of wavelength for the high resolution prism (solid lines) in all three modes (J-, H- and K-band) and the low resolution prism (dashed line) across the entire CHARIS spectral range.}
\label{prismOptions}
\end{figure}

CHARIS uses prisms for dispersion due to their high throughput and large free spectral range. The prism is placed in a collimated beam at the pupil plane. CHARIS will contain two direct vision (or zero deviation) prisms, one for the low and one for the high resolution modes. Both prisms consist of three prism wedges. The design of direct vision prisms is discussed in Nagen, N. \& Tkaczyk, T. S. (2011)\cite{Nagen2011Compound}. The three wedges that make up the high resolution prism are  BaF2, L-BBH1 and a second wedge of BaF2. L-BBH1 is a new Ohara glass optimized for the near infrared and serves as an excellent flint glass when used in a dispersing prism. This pair of glass materials gives relatively even dispersion across the J-, H- and K-bands. A plot of the spectral resolution as a function of wavelength is shown in \fig{prismOptions} for the three high and one low resolution modes. The highest resolution (up to R = 90) is provided in the J-band, whereas the lowest resolution (still at least R = 60) is in the H-band. The average spectral resolution across all wavelengths in the high-res mode is R = 73. The design of the high resolution prism is driven completely by attempting to make all three (J-, H- and K-band) high resolution spectra the same length.

The low resolution prism is designed to disperse light from $1.15\mu m - 2.4\mu m$ simultaneously allowing all three bands to be imaged by CHARIS (at a low resolution) at the same time. The low resolution prism design is still being finalized, but will likely use the same materials as the high-res prism.  Because the low resolution spectra need only create one 20\% bandpass spectrum at a time, the design of the prism was focused on how the light was distributed within the spectra. The design specification for the low resolution mode required that there be at least one spectral measurement in J-band, three measurements in H-band and two in K-band. The reason for having an increased number of spectral measurements at the longer wavelengths is because spectral features in the exoplanet atmospheres are expected to be present here (such as the methane feature in H-band). In addition to the number of resolution elements in each band at least one spectral measurement ({\it i.e.} 2 pixel) gap between each band is necessary to avoid cross-contaminating the bands with each other. 

\section{Transmission \& Noise}\label{sec:TN}
\begin{figure}[h] 
\centering
\includegraphics[width = .6\textwidth]{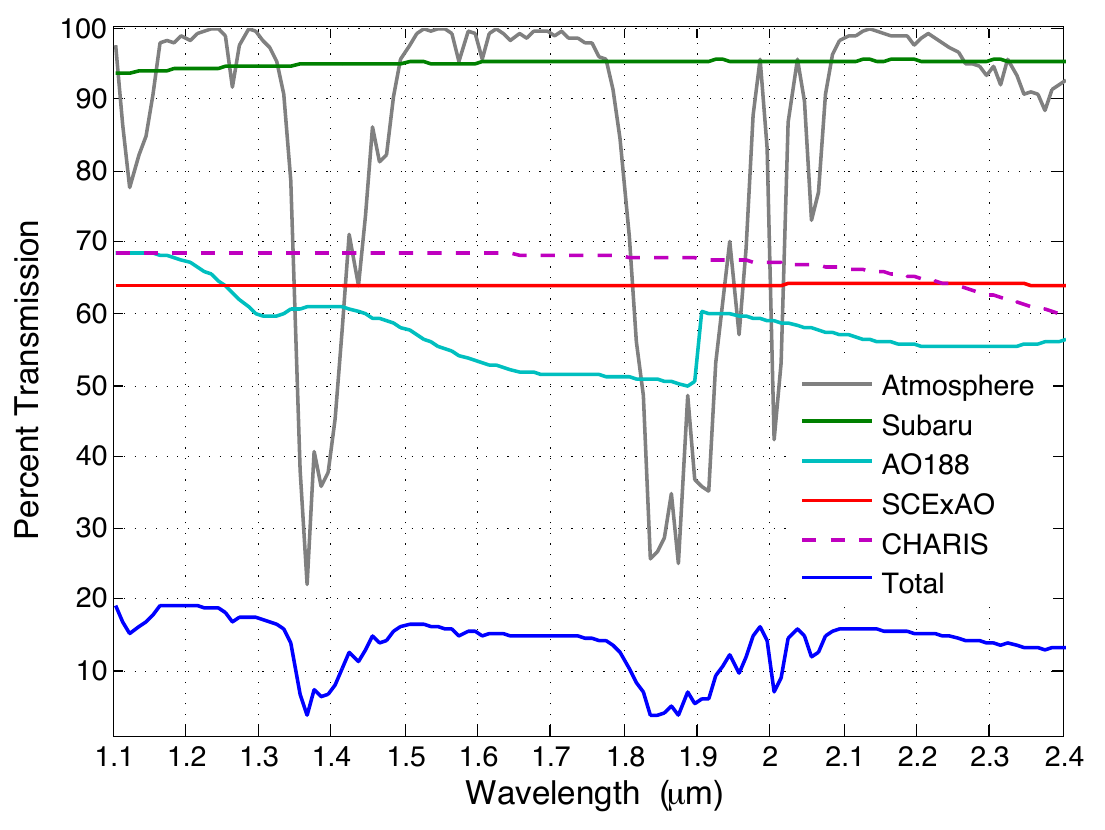}
\caption{Percent transmission as a function of wavelength in high spectral resolution mode for components prior to and including the CHARIS instrument. Quantum efficiency (not shown) is assumed to be 70\% for this calculation. For high res observations, we assume the ADC in the optical path during J-band and H-band observations, but removed during K-band observing -- this is the source of the jump in the AO188 transmission. This plot does not include the CHARIS filter transmission.}
\label{transmission2}
\end{figure}

\fig{transmission2} shows the percent transmission as a function of wavelength in high spectral resolution mode for components prior to and including the CHARIS instrument. Quantum efficiency, which is not shown on the plot, but is included in the calculation of total transmission, is assumed to be 70\% at all wavelengths. The CHARIS transmission is likely slightly optimistic as this represents the design target transmissions, but not the as-built transmission. The filter transmission is also not included. The beam splitter and atmospheric dispersion corrector (ADC) are the main transmission drivers in AO188. We assume the atmospheric dispersion corrector is in the optical path during J-band and H-band observations, but removed during K-band observing -- this is the source of the jump in the AO188 transmission.

\begin{figure}[h] 
\centering
\includegraphics[width = .55\textwidth]{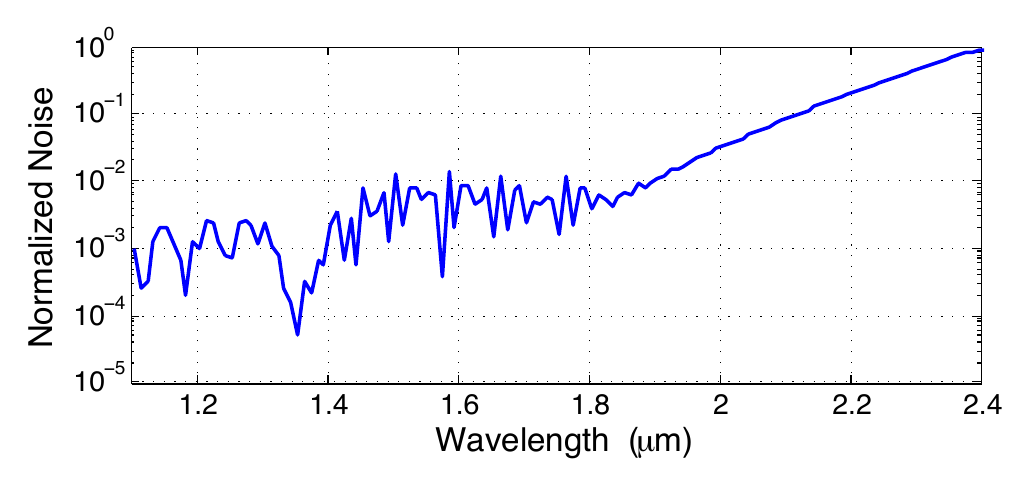}
\caption{Normalized noise at the CHARIS detector as a function of wavelength in high spectral resolution mode for components prior to and including the CHARIS instrument. The noise is dominated by atmosphere emission through the H-band. The noise is $4.2\times$ higher in the H-band than J-band due to the OH lines. The blackbody emission from optics is the dominant noise source in the K-band. Quantum efficiency (not shown) is assumed to be 70\% for this calculation. This plot does not include the CHARIS filter transmission nor the read noise. Note that this noise data are binned to higher resolution than the CHARIS spectra will be to show the features in the noise more clearly.}
\label{NoisePlot}
\end{figure}

\fig{NoisePlot} shows the total normalized noise for the system, including the atmospheric emission and blackbody emission from optics prior to and including CHARIS. The noise is dominated by atmosphere emission in J- and H-band. The noise is $4.2\times$ higher in the H-band than J-band due to the OH lines. The blackbody emission from optics is the dominant source of emission in the K-band. 

\section{CONCLUSIONS}\label{sec:conclusions}

CHARIS will be the first exoplanet-purposed IFS on an 8m class telescope in the northern hemisphere able to achieve a small inner-working angle ($2 \lambda/D$) and high contrasts ($ 10^{-4}-10^{-7} $). CHARIS will provide $R>60$ spectra in J-, H-, and K-bands and low resolution spectra ($R\sim18$) across all three bands simultaneously in a 2.07"$\times$2.07" FOV.  In this paper we presented the optical design in its current state (which is 3 months prior to CDR-level). We presented equations for calculating the fundamental IFS parameters and discussed the details of the CHARIS optical train. 

We also discussed several of the innovative design choices that are unique to the CHARIS IFS. This included the detailed design of the lenslet array and our optimization of incident f-number, lenslet f-number and lenslet pitch which vary significantly (factors of 2--3) from most existing high-constrast, exoplanet-purposed IFSs. We discussed the purpose and layout of the pinholes on the lenslet, which minimize crosstalk for the first time in a exoplanet-purposed IFS. We listed the sources of phase errors (which lead to translation of the PSFlet) on the lenslet. This included a estimation of the residual atmospheric phase errors (post AO correction) which is expected to be the dominant source of phase errors in CHARIS. This paper may represent the first instances in which phase errors on the lenslet were quantified and characterized. We discussed the performance of the spectrograph optics including the motivation for the specified ensquared energy values and their relation to crosstalk. We also noted that the CHARIS spectrograph optics are reflective which minimizes chromaticity and increases throughput -- this is another unique feature in the CHARIS design. The high resolution prism in CHARIS uses a new glass material (L-BBH1) which allows for relatively even resolution across J-, H-, and K-bands enabling a single high-res prism to provide spectra of equal length in the three bands.

Finally, we ended with a calculation of the transmission and noise for the end-to-end system including atmosphere, telescope, both AO systems and, of course, CHARIS. The numerous innovative design features discussed herein will result in higher contrast and will help minimize the noise in CHARIS. The instrument will go on-sky near the end of 2015.

\acknowledgments       
 
This work was performed under a Grant-in-Aid for Scientific Research on Innovative Areas from MEXT of the Japanese government (Number 23103002).



\bibliographystyle{spiebib}

\end{document}